\documentclass[twocolumn]{aastex62}
\usepackage{apjfonts}
\usepackage{lineno}
\usepackage{here}
\accepted{\today}
\submitjournal{ApJ}

\shorttitle{Spatially resolved spectroscopy of thermal X-rays in W44 with {\it XMM-Newton}}
\shortauthors{Okon et al.}

\begin{document}

\title{Deep {\it XMM-Newton} Observations Reveal the Origin of Recombining Plasma in the Supernova Remnant W44}

\correspondingauthor{Hiromichi Okon}
\email{okon@cr.scphys.kyoto-u.ac.jp}

\author{Hiromichi Okon}
\affil{Department of Physics, Kyoto University, Kitashirakawa Oiwake-cho, Sakyo, Kyoto 606-8502, Japan}

\author{Takaaki Tanaka}
\affil{Department of Physics, Kyoto University, Kitashirakawa Oiwake-cho, Sakyo, Kyoto 606-8502, Japan}
	
\author{Hiroyuki Uchida}
\affil{Department of Physics, Kyoto University, Kitashirakawa Oiwake-cho, Sakyo, Kyoto 606-8502, Japan}	

\author{Hiroya Yamaguchi}
\affil{Institute of Space and Astronautical Science, JAXA, 3-1-1 Yoshinodai, Chuo, Sagamihara, Kanagawa 252-5210, Japan}

\author{Takeshi Go Tsuru}
\affil{Department of Physics, Kyoto University, Kitashirakawa Oiwake-cho, Sakyo, Kyoto 606-8502, Japan}

\author{Masumichi Seta}
\affil{Department of Physics, School of Science and Technology, Kwansei Gakuin University, 2-1 Gakuen, Sanda, Hyogo 669-1337, Japan}

\author{Randall K. Smith}
\affil{Harvard-Smithsonian Center for Astrophysics, 60 Garden St, Cambridge, MA 02138, USA}

\author{Satoshi Yoshiike}
\affil{Department of Physics, Nagoya University, Furo-cho, Chikusa, Nagoya, Aichi 464-8601, Japan}

\author{Salvatore Orlando}
\affil{INAF-Osservatorio Astronomico di Palermo, Piazza del Parlamento 1, 90134 Palermo, Italy}

\author{Fabrizio Bocchino}
\affil{INAF-Osservatorio Astronomico di Palermo, Piazza del Parlamento 1, 90134 Palermo, Italy}

\author{Marco Miceli}
\affil{Dipartimento di Fisica e Chimica E. Segre, Universita degli studi di Palermo, Piazza del Parlamento 1, 90134 Palermo, Italy}
\affil{INAF-Osservatorio Astronomico di Palermo, Piazza del Parlamento 1, 90134 Palermo, Italy}

%
%
%



\begin{abstract}
Recent X-ray studies revealed over-ionized recombining plasmas (RPs) in a dozen mixed-morphology (MM) supernova remnants (SNRs).
However, the physical process of the over-ionization has not been fully understood yet.
Here we report on spatially resolved spectroscopy of X-ray emission from W44, one of the over-ionized MM-SNRs, 
using {\it XMM-Newton} data from deep observations, aiming to clarify the physical origin of the over-ionization. 
We find that combination of low electron temperature and low recombination timescale is achieved in the region interacting with dense molecular clouds.
Moreover, a clear anti-correlation between the electron temperature and the recombining timescale is obtained from each of the regions with and without the molecular clouds. 
The results are well explained if the plasma was over-ionized by rapid cooling through thermal conduction with the dense clouds hit by the blast wave of W44. 
Given that a few other over-ionized SNRs show evidence for adiabatic expansion as the major driver of the rapid cooling, 
our new result indicates that both processes can contribute to over-ionization in SNRs, with the dominant channel depending on the evolutionary stage.  
\end{abstract}



\section{Introduction} 

In the previously accepted scenario, plasmas in supernova remnants (SNRs) were believed to 
be collisionally ionizing until they reach an equilibrium. 
Thus, the ionization degree was expected to be equal to or below that in collisional ionization equilibrium (CIE).  
The standard scenario, however, was questioned by \cite{Kawasaki2002}. 
Analyzing {\it ASCA} data of IC~443, 
they measured the ionization degree of S ions based on a Ly$\alpha$-to-He$\alpha$ line intensity ratio implying an ionization degree 
higher than that in the CIE. 
{\it Suzaku} data then provided clearer evidence of over-ionization. 
\cite{Ozawa2009} and \cite{Yamaguchi2009} discovered strong radiative recombining continua (RRCs) in {\it Suzaku} data of W49B and IC~443, respectively, 
directly indicating that the plasmas are in a recombination-dominant state. 

Over-ionized recombining plasmas (RPs) have since been discovered in a dozen SNRs \citep{Uchida2015}.
All of these SNRs are classified as mixed-morphology (MM) SNRs, which are characterized by radio shells  
with center-filled X-ray emissions \citep{Rho1998}.
Most of the MM-SNRs are known to be interacting with molecular clouds.
These facts imply that the presence of the ``recombination dominant phase''  is somewhat common among MM SNRs, and that 
shock-cloud interactions seem to be related to the origin of RPs. 
Based on spatially-resolved spectroscopy of {\it Suzaku} data, some authors such as \cite{Matsumura2017b} and \cite{Okon2018} 
claimed that X-ray emitting plasmas in, e.g., IC~443 and W28 were rapidly cooled due to thermal conduction with 
interacting molecular clouds, which resulted in the formation of the RPs as originally proposed by \cite{Kawasaki2002} and  \cite{Kawasaki2005}. 
\cite{Itoh1989} and \cite{Shimizu2012} predicted another scenario, the so-called rarefaction scenario, in which rapid adiabatic expansion is responsible for the over-ionization. 
Some authors, e.g.,  \cite{Miceli2010}, \cite{Lopez2013}, \cite{Greco2018}, and \cite{Sezer2019}, indeed claimed 
that their X-ray spectroscopy results support the scenario. 
The clearest evidence was presented by  \cite{Yamaguchi2018}, who analyzed {\it NuSTAR} data of W49B and performed spatially resolved spectroscopy, 
focusing on the Fe RRC. 
With these observational results contradicting each other, the physical origin of RPs is not understood yet and is still under debate.

\begin{figure*}[ht]
\begin{center}
\vspace{-0mm}
\includegraphics[width=16cm]{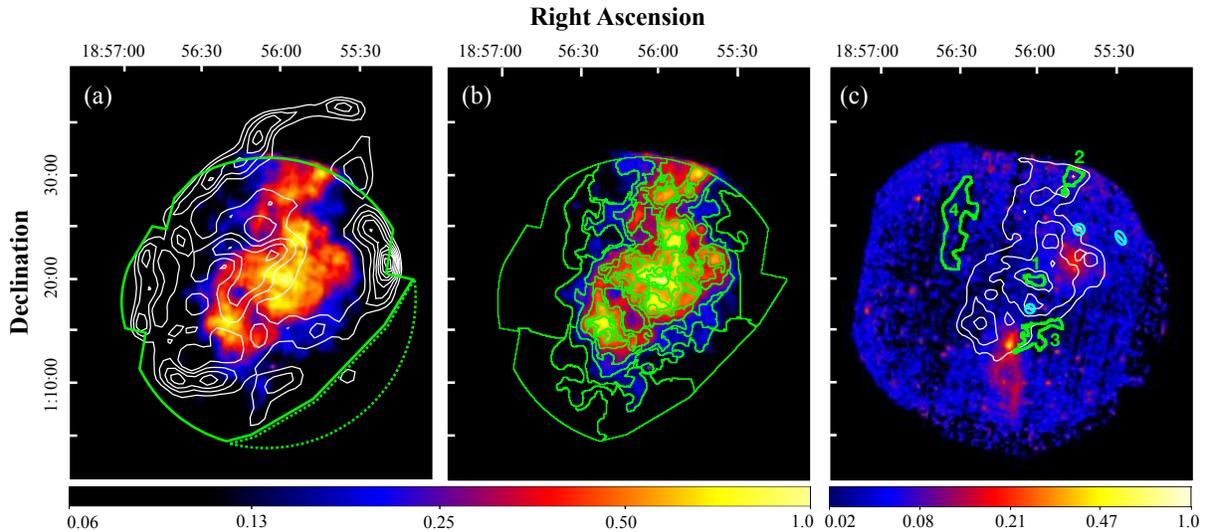} 
\end{center}
\vspace{-2mm}
\caption{
MOS$+$pn images of W44 in the energy band of (a)(b) 0.5--4.0~keV and (c) 4.0--8.0~keV after NXB subtraction and correction for the vignetting effect. 
The coordinate refers to the J2000.0 epoch. 
The white contours in panel (a) indicate a radio continuum image at 1.4 GHz taken with the Karl G. Jansky Very Large Array whereas 
those in pane (c) are the 0.5--4.0 keV X-ray image. 
The source and background spectra were extracted from the regions enclosed by the solid and dashed lines in panel (a). 
The source region was divided into 70 sub-regions as shown in panel (b). 
The regions enclosed by the green lines in panel (c) are the four representative sub-regions whose spectra are plotted in Figure~\ref{fig:W44_spectrum}. 
The cyan ellipses are regions excluded in the spectral analysis to remove bright point sources. 
}
\label{fig:Xray_image}
\end{figure*}

W44 (a.k.a., G34.7$-$0.4 or 3C~392) is a Galactic MM SNR with an estimated distance of $\sim 3$~kpc \citep{Claussen1997,Ranasinghe2018}.
The age of W44 is estimated to be $\sim 20~{\rm kyr}$ \citep{Smith1985a,Wolszczan1991,Harrus1997}. 
\cite{Wolszczan1991} discovered a radio pulsar, PSR B1853$+$01, in the southern part of the remnant, indicating that W44 is a remnant of a core-collapse supernova.
W44 is known to be interacting with molecular clouds as evidenced by radio observations of OH masers 1720~MHz \citep{Frail1998,Claussen1999} and $^{12}$CO lines \citep{Seta1998,Seta2004,Yoshiike2013}.
In the X-ray band, \cite{Jones1993} observed W44 with {\it Einstein} and reported that the emission is predominantly thermal, based on the presence of Mg, Si, and S emission lines.
In {\it Suzaku} data, \cite{Uchida2012} found RRCs of Si and S, and concluded that the plasmas in the central bright region (Figure~\ref{fig:Xray_image}) are in an over-ionized state.

Here we report on results from a spatially resolved analysis of {\it XMM-Newton} data of deep observations of W44 
to pin down the physical origin of the RP.
Throughout the paper, errors are quoted at 90\% confidence levels in the tables and text. 
Error bars shown in figures correspond $1 \sigma$ confidence levels.

\section{Observations}

\begin{deluxetable*}{lcccc}
\tablecaption{Observation log. \label{tab:obs_id}}
\tablecolumns{5}
\tablewidth{0pt}
\tablehead{
\colhead{Target} &
\colhead{Obs. ID} &
\colhead{Obs. Date} &
\colhead{(R.A., Dec.)\tablenotemark{a}} &
\colhead{Effective Exposure}
}
\startdata
	W44~PWN & 0551060101 & 2009 April 24  & ($18^{\rm h} 56^{\rm m} 11\fs00,~+01\degr 13\arcmin 28\farcs0$) & 65~ks \\
     	W44 & 0721630101 & 2013 October 18 & ($18^{\rm h} 56^{\rm m} 06\fs99,~+01\degr 17\arcmin 54\farcs0$) & 110~ks \\
	W44 & 0721630201 & 2013 October 19 & ($18^{\rm h} 56^{\rm m} 06\fs99,~+01\degr 17\arcmin 54\farcs0$) & 92~ks \\
	W44 & 0721630301 & 2013 October 23 & ($18^{\rm h} 56^{\rm m} 06\fs99,~+01\degr 17\arcmin 54\farcs0$) & 93~ks \\
\enddata
\tablenotetext{a}{Equinox in J2000.0.}
\end{deluxetable*}

W44 was observed several times from 2003 to 2013 with {\it XMM-Newton}. 
We discarded datasets whose effective exposures are extremely short because of flaring backgrounds, 
As a result, four datasets (Obs.ID=0551060101, 0721630101, 0721630201, and 0721630301) were left for further analysis.  
The details of the observations used are summarized in table \ref{tab:obs_id}.
In what follows, we analyze data obtained with the European Photon Imaging Camera (EPIC), which is composed of two 
MOS \citep{Turner2001} and one pn \citep{Strder2001} CCD cameras. 

Following the cookbook for analysis procedures of extended sources\footnote{ftp://xmm.esac.esa.int/pub/xmm-esas/xmm-esas.pdf}, 
we reduced the data with the Science Analysis System software version 16.0.0 and the calibration database version 3.9 released in January 2, 2017\footnote{http://xmm2.esac.esa.int/docs/documents/CAL-TN-0018.pdf}. 
We estimated the non-X-ray background (NXB) with {\tt mos-back}.
We generated the redistribution matrix files and the ancillary response files by using {\tt mos-spectra}. 
We used version 12.9.0u of the XSPEC software \citep{Arnaud1996} for the following spectral analysis.
In the image analysis, we merged MOS1, MOS2 and pn data of the each observation for better photon statistics.
In the spectral analysis, we only used the MOS data because of their lower detector background level than the pn data.

\section{Analysis and Results} \label{sec:floats}
\subsection{Imaging Analysis}
Figure~\ref{fig:Xray_image} shows vignetting- and exposure-corrected images of W44 taken with the EPIC after NXB subtraction. 
The soft-band image in Figure~\ref{fig:Xray_image}(a) reveals the center-filled morphology and small bright knots as already reported by \cite{Shelton2004} based on {\it Chandra} data.
In order to perform a spatially-resolved spectroscopic analysis, we applied the contour binning algorithm \citep{Sanders2006} to the 0.5--4.0 keV image, and divided the source region in Figure~\ref{fig:Xray_image}(a) into 70 subregions as displayed in Figure \ref{fig:Xray_image}(b).
The algorithm generates subregions along the structure of the surface brightness so that each subregion has almost the same signal-to-noise ratio.
We manually excluded bright point sources identified in the hard band image in Figure \ref{fig:Xray_image}(c).

\subsection{Background Estimation}\label{background}
To estimate the X-ray background, 
we extracted spectra from the off-source region (Figure \ref{fig:Xray_image}) in the field-of-views of each observation. 
After subtracting the NXB from each of the spectra, we co-added them to perform spectral fitting. 
The applied model consists of the cosmic X-ray background (CXB), the Galactic ridge X-ray emission (GRXE), and Al and Si K$\alpha$ lines of instrumental origin which are not included in the NXB spectra estimated with {\tt mos-back} \citep{Lumb2002}.
By referring to \cite{Kushino2002}, the CXB component was expressed as a power law with a photon index of $1.4$ and a 2--10~keV intensity of $6.38\times10^8~{\rm erg}~{\rm cm}^{-2}~{\rm s}^{-1}~{\rm sr}^{-1}$. 
We employed the GRXE model by \cite{Uchiyama2013}, which is composed of the foreground emission (FE), the high-temperature plasma emission (HP), the low-temperature plasma emission (LP), and the emission from cold matter (CM). 
We used the Tuebingen-Boulder interstellar medium (ISM) absorption model \citep[TBabs;][]{Wilms2000} to estimate the column density (${N_{\rm H}}^{\rm GRXE}$) for the total Galactic absorption in the line of sight toward W44.
Most of the parameters of GRXE were fixed to those shown by \cite{Uchiyama2013}.
Free parameters are ${N_{\rm H}}^{\rm GRXE}$, $kT_e$ of the LP, and the normalization of each component.
The normalization of \ion{Al}{1} and \ion{Si}{1} K$\alpha$ lines were allowed to vary.
The best-fit parameters are summarized in table \ref{tab:bkg_model}.
We used the best-fit model to account for the X-ray background in the source spectra in \S\ref{subsec:spec}.

\begin{deluxetable*}{cccc}[ht]
\tablecaption{Best-fit model parameters of the background spectrum.\label{tab:bkg_model}}
\tablehead{
\colhead{Component} &
\colhead{Model function} & 
\colhead{Parameter} &
\colhead{Value}
} 
\startdata
      FE & TBabs (Absorption) & ${N_{\rm H}}^{\rm FE}$ (10$^{22}$ cm$^{-2}$) & 0.56 (fixed) \\
      & APEC (FE$_{\rm low}$) & $kT_e$ (keV) & 0.09 (fixed) \\
      & & $Z_{\rm all}$ (solar) &  0.05 (fixed) \\
      & & Norm\tablenotemark{a} &  1.05$^{+0.12}_{-0.16}$ \\
      & APEC (FE$_{\rm high}$) & $kT_e$ (keV) & 0.59 (fixed) \\
      & & $Z_{\rm all}$ &  0.05 (fixed)\\
      & & Norm\tablenotemark{a} &  $\leq$ 1.68 $\times$ 10$^{-3}$ \\ \hline
      GRXE & TBabs (Absorption) & ${N_{\rm H}}^{\rm GRXE}$ (10$^{22}$ cm$^{-2}$) & 1.71$^{+0.10}_{-0.03}$ \\
      & APEC (LP) & $kT_e$ (keV) & 0.56$^{+0.03}_{-0.03}$ \\
      & & $Z_{\rm Ar}$ &  1.07 (fixed)\\
      & & $Z_{\rm other}$ &  0.81 (fixed)\\
      & & Norm\tablenotemark{a} &  1.53$^{+0.02}_{-0.01}\times$ 10$^{-2}$ \\
      & APEC (HP) & $kT_e$ (keV) & 6.64 (fixed) \\
      & & $Z_{\rm Ar}$ &  1.07 (fixed)\\
      & & $Z_{\rm other}$ &  0.81 (fixed)\\
      & & Norm\tablenotemark{a} & = LP Norm. $\times$ 0.29   \\
      & Power law (CM) & $\Gamma$ & 2.13 (fixed) \\
      & & Norm\tablenotemark{b} &  $\leq$ 0.20 \\
     &  Gauss (CM; \ion{Fe}{1} K$\alpha$) & Equivalent Width (eV) & 457 (fixed) \\ \hline
      CXB & TBabs (Absorption) & ${N_{\rm H}}^{\rm CXB}$ & = ${N_{\rm H}}^{\rm GRXE}\times2$\\
      & Power law & $\Gamma$ & 1.40 (fixed) \\
      & & Norm\tablenotemark{b} &  9.69 (fixed)  \\
      \hline
       & & {$\chi^{2}_{\nu}$ ($\nu$)}\tablenotemark{c} & 1.64 (229) \\
\enddata
\tablenotetext{a}{The unit is photons s$^{-1}$ cm$^{-2}$ keV$^{-1}$ sr$^{-1}$ at 1~keV.}
\tablenotetext{b}{The emission measure integrated over the line of sight, i.e., $(1/4\pi D^2) \int n_e n_{\rm H} dl$ in the unit of $10^{-14}$~cm$^{-5}$ ~sr$^{-1}$.}
\tablenotetext{c}{The parameters ${\chi_\nu}^2$ and $\nu$ indicate a reduced chi-squared and a degree of freedom, respectively.}
\end{deluxetable*}

\begin{figure}[ht]
\begin{center}
 \includegraphics[width=8.0cm]{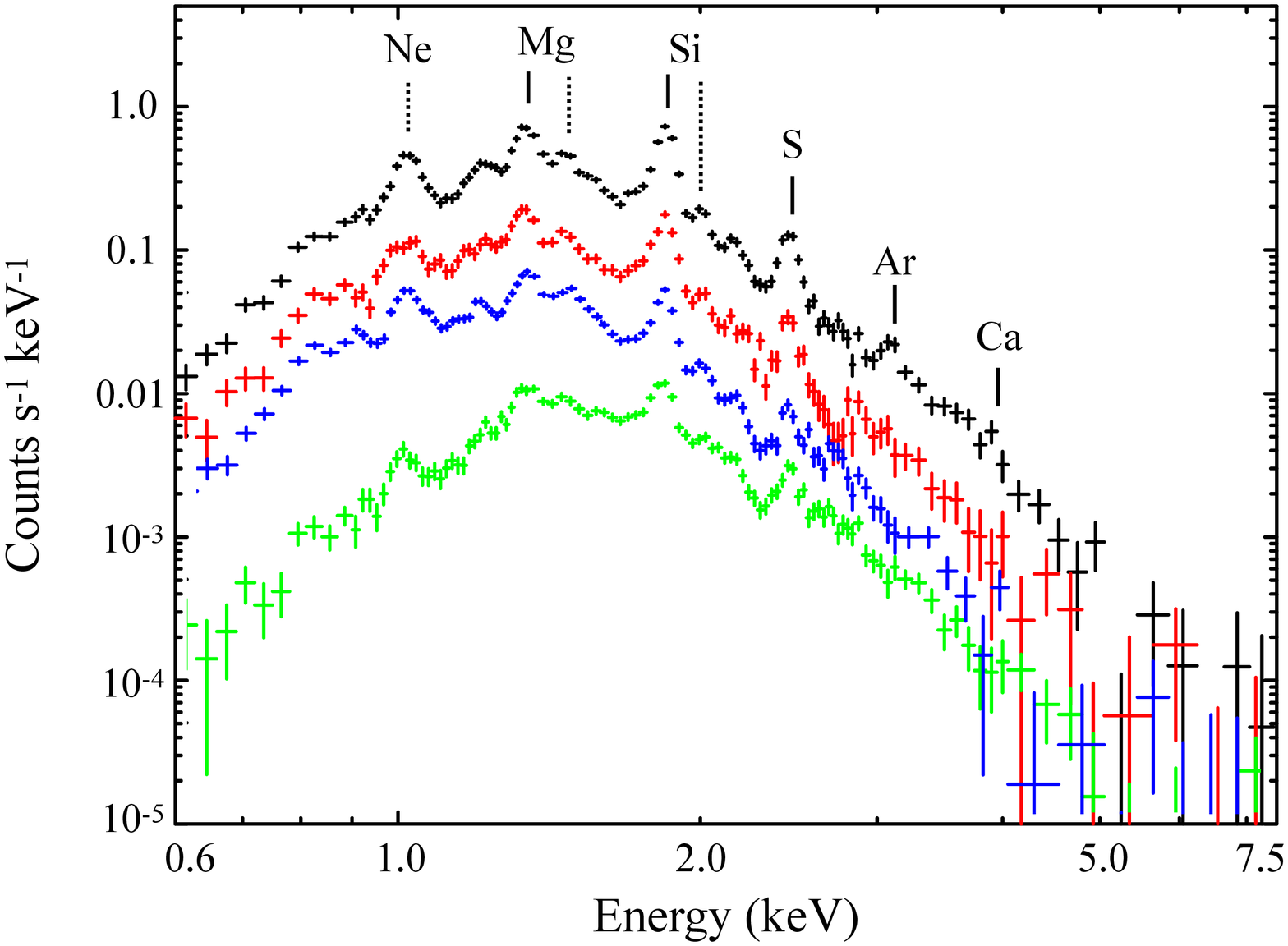} 
\end{center}
\vspace{0mm}
\caption{
MOS (MOS1 + MOS2) spectra extracted from the sub-regions 1 (black), 2 (red), 3 (blue), and 4 (green), whose locations are shown in Figure \ref{fig:Xray_image}(c). 
The NXB and X-ray background are subtracted. 
For a display purpose, the spectra of Regions 3 and 4 are scaled by factors of 0.05 and 0.1, respectively. 
The vertical solid and dashed lines denote the centroid energies of the He$\alpha$ lines and Ly$\alpha$ lines, respectively. 
}
\label{fig:W44_spectrum}
\end{figure}

\subsection{Spectral Analysis}\label{subsec:spec}
\begin{figure}
\vspace{2mm}
\begin{center}
 \includegraphics[width=6cm]{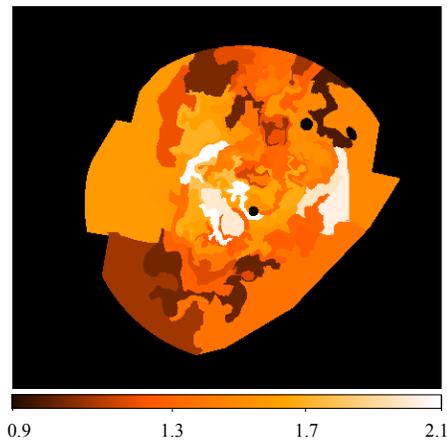} 
\end{center}
\vspace{-3mm}
\caption{
Reduced chi-squared map. 
}
\label{fig:chi}
\end{figure}
\begin{figure*}
\begin{tabular}{cccc}
\begin{minipage}[c]{0.5\hsize}
 \includegraphics[width=8cm]{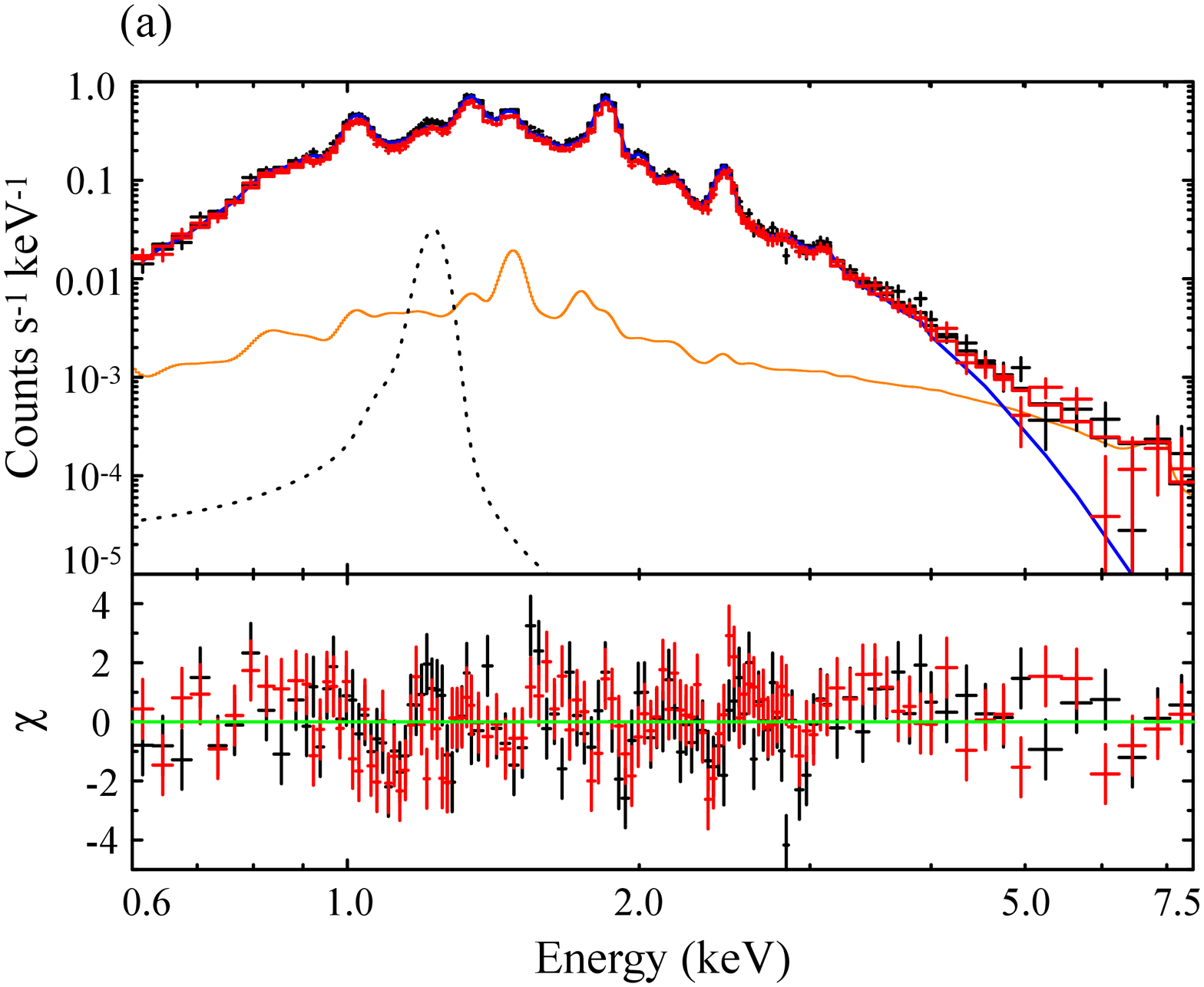} 
\end{minipage}
\begin{minipage}[c]{0.5\hsize}
 \includegraphics[width=8cm]{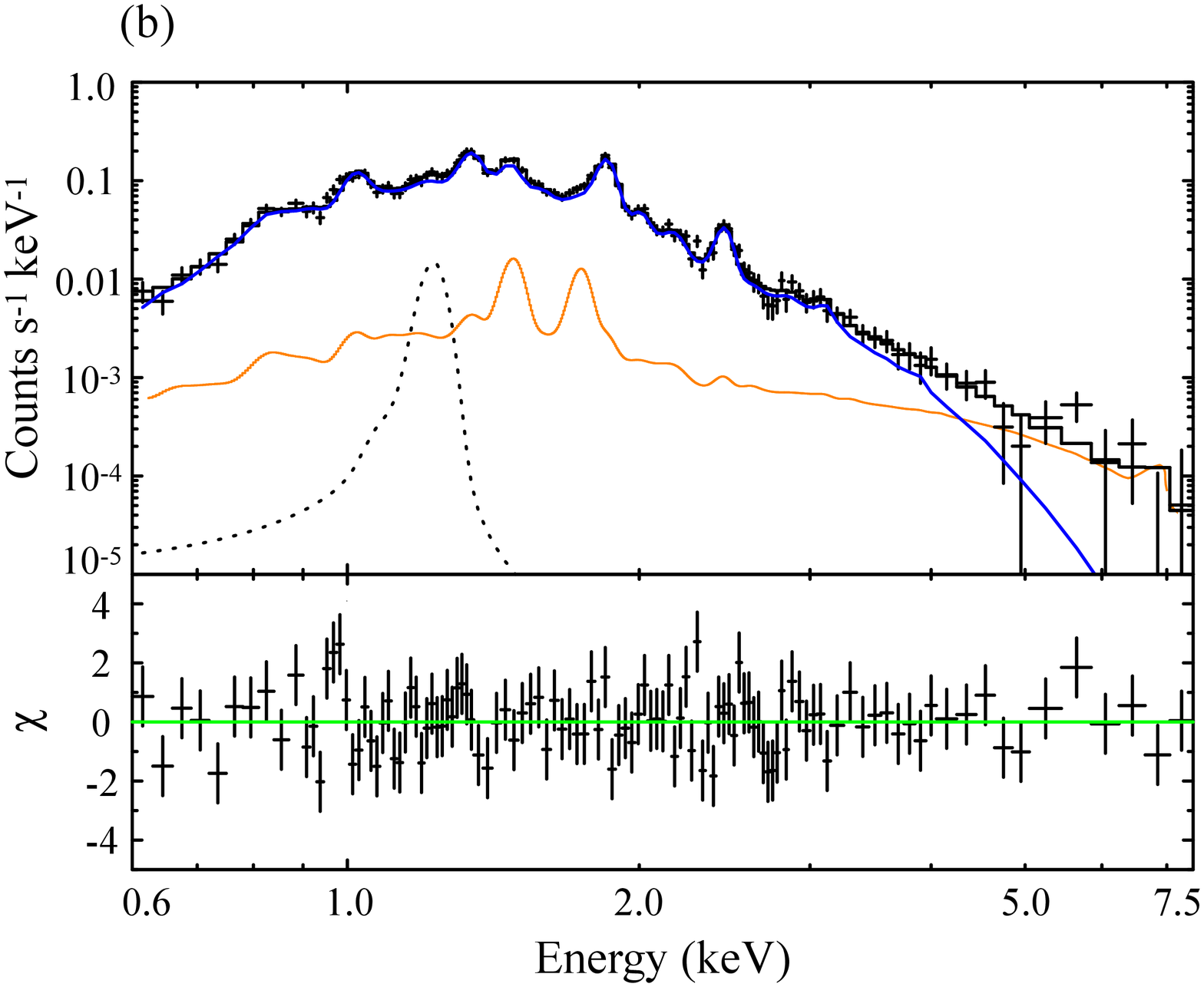} 
\end{minipage}
\\\\
\begin{minipage}[c]{0.5\hsize}
\vspace{-5mm}
 \includegraphics[width=8cm]{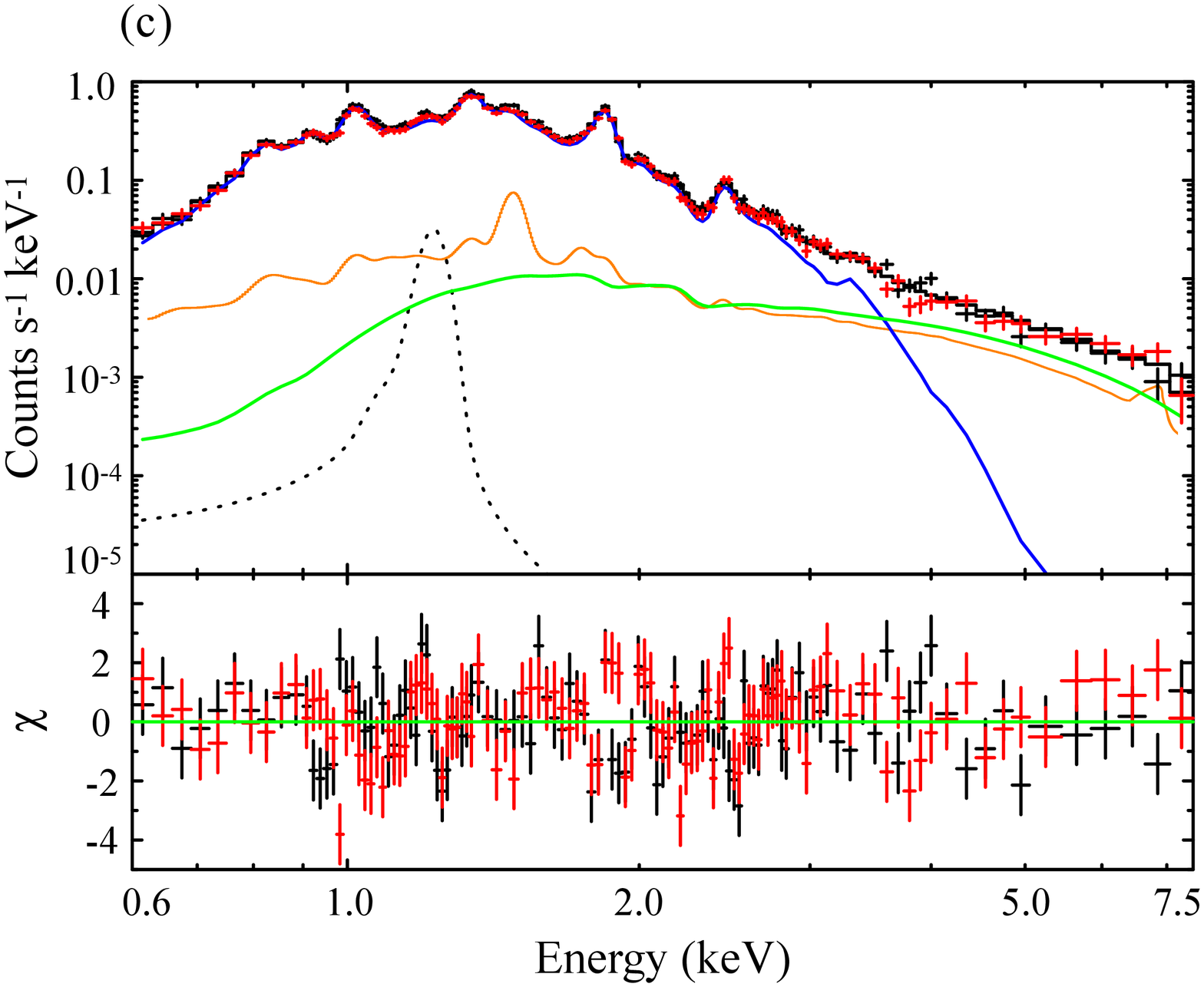} 
\end{minipage}
\begin{minipage}[c]{0.5\hsize}
\vspace{-5mm}
 \includegraphics[width=8cm]{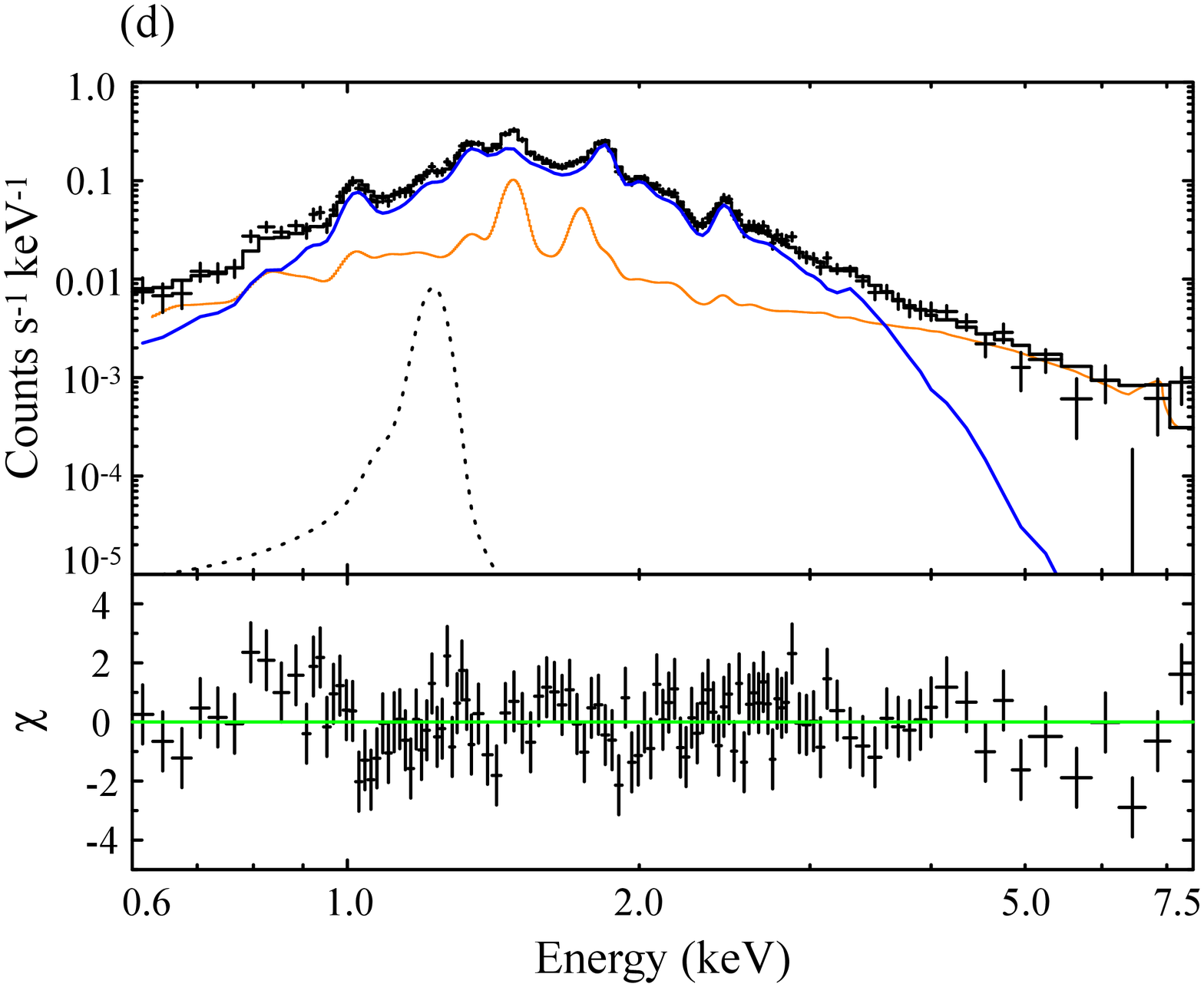} 
\end{minipage}
\end{tabular}
\vspace{0mm}
\caption{
(a) MOS1 (red) and MOS2 (black) spectra from Region~1 plotted with the model without a Gaussian for Fe L lines (see text for details). 
The blue, orange, and black curves represent the RP model, the background, and the sum of the models, respectively. 
The black dotted curve represent a Gaussian at 1.23 keV added to the model (the black dotted curve). 
(b)--(d) Same as panel (a) but for Regions~2--4. 
Since Region~3 is contaminated by the southern hard source (PSR B1853$+$01 and its PWN), the model includes a power law shown as the green curve. 
Only MOS2 data are available in Regions 2 and 4 because of malfunction of CCD chips of MOS1. 
}
\label{fig:spectrum1}
\end{figure*}

\begin{deluxetable*}{cccccc}[ht]
\tablecaption{Best-fit model parameters of the spectra from the representative sub-regions.\label{tab:plasma_model}}
\tablehead{
\colhead{Model function} &
\colhead{Parameters} &
\colhead{Region 1} & 
\colhead{Region 2} &
\colhead{Region 3} &
\colhead{Region 4}
} 
\startdata
      TBabs & $N_{\rm H}$ (10$^{22}$ cm$^{-2}$) & 1.51$^{+0.01}_{-0.03}$ & 1.43$\pm$0.07 & 1.83$\pm$0.03 & 2.8$\pm$0.1 \\
      VVRNEI & $kT_e$ (keV) & $0.503^{+0.019}_{-0.003}$ & 0.55$^{+0.04}_{0.03}$& 0.239$^{+0.004}_{-0.002}$ & 0.26$\pm$0.01\\
      & $kT_{\rm init}$ (keV) & 1.0 (fixed) & 1.0 (fixed) & 1.0 (fixed) & 1.0 (fixed) \\
      & $Z_{\rm Ne}$ (Solar) & 1.29$^{+0.06}_{-0.07}$ & 0.8$\pm$0.1 & 1.7$\pm$0.1 & 3.0$\pm$1.0 \\
      & $Z_{\rm Mg}$ (Solar) & 1.55$^{+0.07}_{-0.05}$ & 1.1$\pm$0.1 & 1.5$\pm$0.1 & 2.2$^{+0.8}_{-0.4}$ \\
      & $Z_{\rm Si}$ (Solar) & 2.68$^{+0.09}_{-0.05}$ & 1.7$\pm$0.1 & 2.6$^{+0.2}_{-0.1}$ & 2.6$^{+0.9}_{-0.5}$ \\
      & $Z_{\rm S}=Z_{\rm Ar}=Z_{\rm Ca}$ (Solar) & 2.1$\pm$0.1 & 1.3$\pm$0.2 & 3.1$^{+0.4}_{-0.2}$ & 3.2$\pm$1.0 \\ 
      & $Z_{\rm Fe}=Z_{\rm Ni}$ (Solar) & 0.14$^{+0.02}_{-0.01}$ &0.17$^{+0.05}_{-0.04}$ & 1.1$\pm$0.1 & 1.0$^{+0.6}_{-0.5}$ \\
      & $n_et$ (10$^{11}$ cm$^{-3}$s) & 5.3$^{+0.2}_{-0.1}$ & 6.0$\pm$0.1 & 6.0$\pm$0.2 & 4.0$\pm$0.3 \\
      & Norm\tablenotemark{a} & 0.048$^{+0.005}_{-0.003}$ & 0.026$^{+0.006}_{-0.005}$ & 0.11$\pm0.04$ & 0.19$^{+0.02}_{-0.01}$ \\ \hline
      {$\chi^{2}_{\nu}$ ($\nu$)}\tablenotemark{b} & & 1.45 (217) & 1.12 (104) & 1.57 (215) & 1.25 (104) \\
\enddata
\tablenotetext{a}{The emission measure integrated over the line of sight, i.e., $(1/4\pi D^2) \int n_e n_{\rm H} dl$ in the unit of $10^{-14}$~cm$^{-5}$ ~sr$^{-1}$.}
\tablenotetext{b}{The parameters ${\chi_\nu}^2$ and $\nu$ indicate a reduced chi-squared and a degree of freedom, respectively.}
\end{deluxetable*}

Figure \ref{fig:W44_spectrum} shows background-subtracted MOS spectra extracted from the representative four sub-regions shown in Figure \ref{fig:Xray_image}(c), where emission lines 
from highly ionized Ne, Mg, Si, S, Ar and Ca are clearly resolved. 
The Ly$\alpha$-to-He$\alpha$ ratios and the continuum shape below $\sim 2$~keV are different from each other, suggesting significant region-to-region variations of plasma parameters and of absorption column densities. 
We first fitted spectra from all 70 sub-regions with a CIE model.  
All the fittings left hump-like residuals at $\sim 2.7~{\rm keV}$ corresponding to the edge of the \ion{Si}{13} RRC. 
The result implies that the X-ray emitting plasma in W44 is over-ionized not only in a part of the 
remnant, as reported by \cite{Uchida2012} based on {\it Suzaku} data of the central region, but in the whole remnant. 



We then fitted all spectra with the RP model, VVRNEI \citep{Foster2017}, in XSPEC.
The model describes emission from a thermal plasma $n_et$ after an abrupt decrease of the electron temperature from $kT_{\rm init}$ to $kT_e$ under an assumption that the plasma 
initially was in CIE. 
Using the TBabs model, we took into account photoelectric absorption by the foreground gas with the solar abundances of \cite{Wilms2000}. 
We allowed the column density $N_{\rm H}$,  the electron temperature $kT_e$, recombining timescale $n_et$, and normalization of the VVRNEI component to vary. 
The parameter $kT_{\rm init}$ was fixed in the fittings. 
We tried $kT_{\rm init}$ of 1.0~keV, 2.0~keV, 3.0~keV, and 5.0~keV, and found that the data are best reproduced with $kT_{\rm init} = 1.0~{\rm keV}$. 
We thus show results obtained with $kT_{\rm init}$ fixed at 1.0~keV in what follows. 
We note that parameters such as $N_{\rm H}$, $kT_e$, and $n_et$ are insensitive to the choice of $kT_{\rm init}$. 
The abundances of Ne, Mg, Si, S, and Fe were left free, whereas Ar and Ca were linked to S, and Ni was linked to Fe. 
The other abundances were fixed to the solar values. 
To the model for the W44 emission, we added the background model with the parameters fixed to those in Table~\ref{tab:bkg_model}. 
We allowed the normalizations of the Al and Si K$\alpha$ lines to vary since the line intensities are known to have location-to-location variations on the detector plane \citep{Kuntz2008}.

In the fittings, we noticed line-like residuals at $\sim 1.2$~keV in most of the sub-regions.
Previous studies pointed out that the residuals are most probably due to the lack of Fe-L lines (e.g., \ion{Fe}{17} n > 6 $\rightarrow$ n = 2) in the atomic code \cite[e.g., ][]{Matsumura2017b}.
However, the used NEI plasma code takes into account of the Fe-L lines \citep{Foster2017}.
The residuals could be attributed to uncertainty in emissivity data of Fe-L lines in the code or to physical processes such as charge exchange that are not taken into account here.
While the reason for the residuals is not clear, 
an addition of a Gaussian at 1.23 keV significantly improved the fits. 
In the case of Region 1 shown in Figure~\ref{fig:spectrum1}(a), the fitting statistic was improved from ${\chi_\nu^2}$ = 1.54 with $\nu$ = 218 to ${\chi_\nu^2}$ = 1.45 with $\nu$ = 217. 
On the other hand, the addition of the Gaussian does not change the parameters obtained beyond the 90\% confidence level.
In the spectral analyses, therefore, we used the model including the Gaussian.

Additional components are necessary to fit the spectra from the sub-regions where two extended hard sources are detected (Figure \ref{fig:Xray_image}(c)). 
\cite{Nobukawa2018} pointed out that the northeastern source, which was discovered by \cite{Uchida2012}, is probably a galaxy cluster. 
To account for the emission, therefore, we employed a CIE model, in which the abundance and $kT_e$ were fixed to the values as determined 
by \cite{Matsumura2018} while the normalization was left free.
The southern source encompasses PSR B1853$+$01 and its associated pulsar wind nebular (PWN) detected with {\it Chandra} \citep{Petre2002}. 
Since their X-ray emissions have featureless continuum spectra \citep{Petre2002}, we used a power law model for the southern source, in which the photon index $\Gamma$ and the normalization 
were free parameters. 

All 70 spectra are well reproduced by the above RP (with an additional component) model.
In addition to the result from Region 1 in Figure~\ref{fig:spectrum1}(a), we plot the best-fit models overlaid on the observed spectra from Regions 2--4 in Figure~\ref{fig:spectrum1}(b)--(d). 
In a map in Figure~\ref{fig:chi}, we present ${\chi_\nu^2}$ values of each sub-region, which range from 0.98 to 2.15. 
Among all 70 sub-regions, 20 sub-regions have ${\chi_\nu^2} \geq$~1.5, and
4 sub-regions have ${\chi_\nu^2} \geq$ 2.0.

\section{Discussion}

\subsection{Foreground Gas Distribution}

\begin{figure*}[ht]
\begin{center}
 \includegraphics[width=10cm]{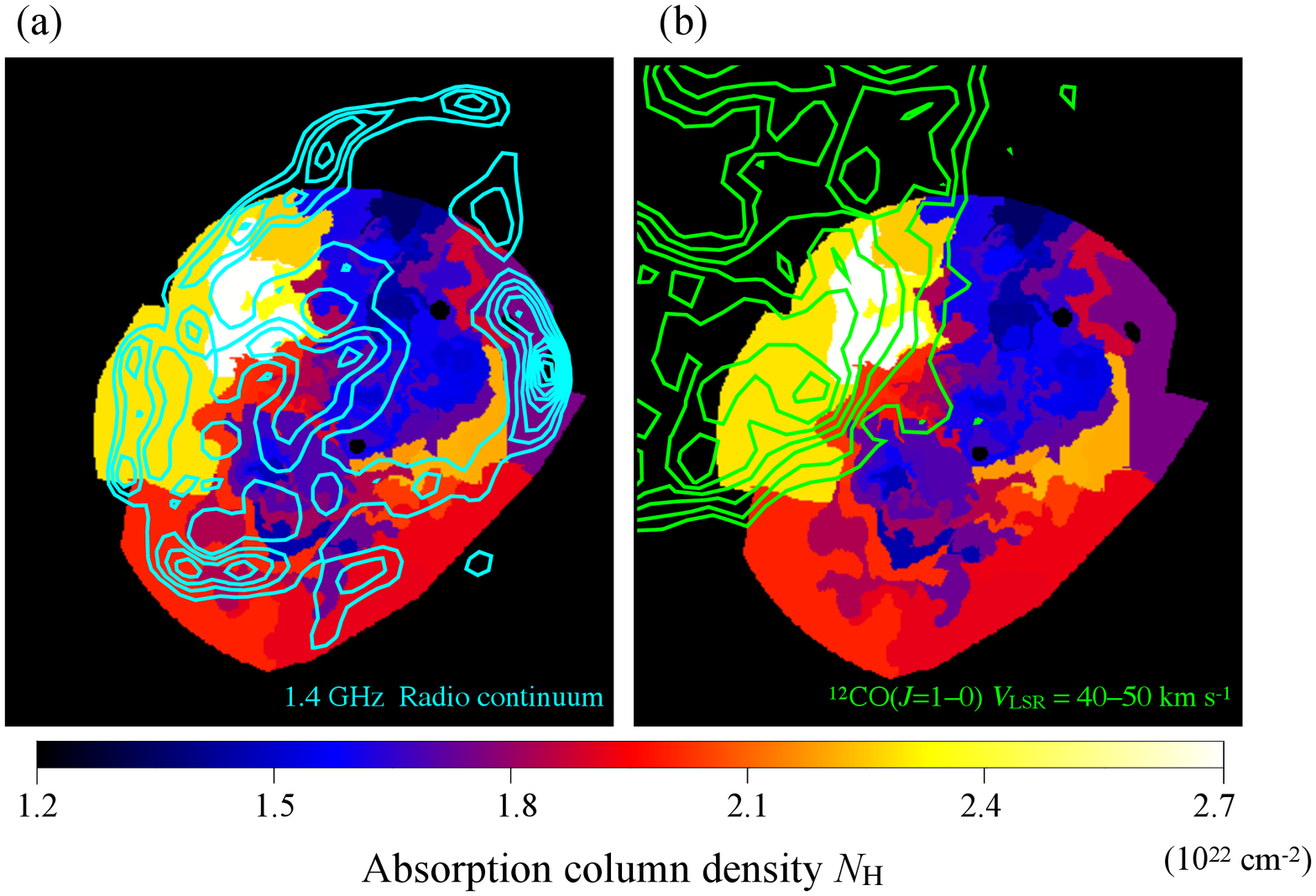} 
\end{center}
\vspace{-4mm}
\caption{
Distribution of X-ray absorption column density ($N_{\rm H}$). 
The same radio continuum image as that in Figure~\ref{fig:Xray_image}(a) are overlaid as cyan contours in panel (a). 
The green contours in panel (b) denote $^{12}$CO($J=1$--$0$) emissions in a velocity range of $V_{\rm LSR} = 40$--$50~{\rm km}~{\rm s}^{-1}$ as observed with the Nobeyama 45 m radio telescope in the FUGIN project \citep{Umemoto2017}. 
}
\label{fig:nH}
\begin{center}
 \includegraphics[width=15cm]{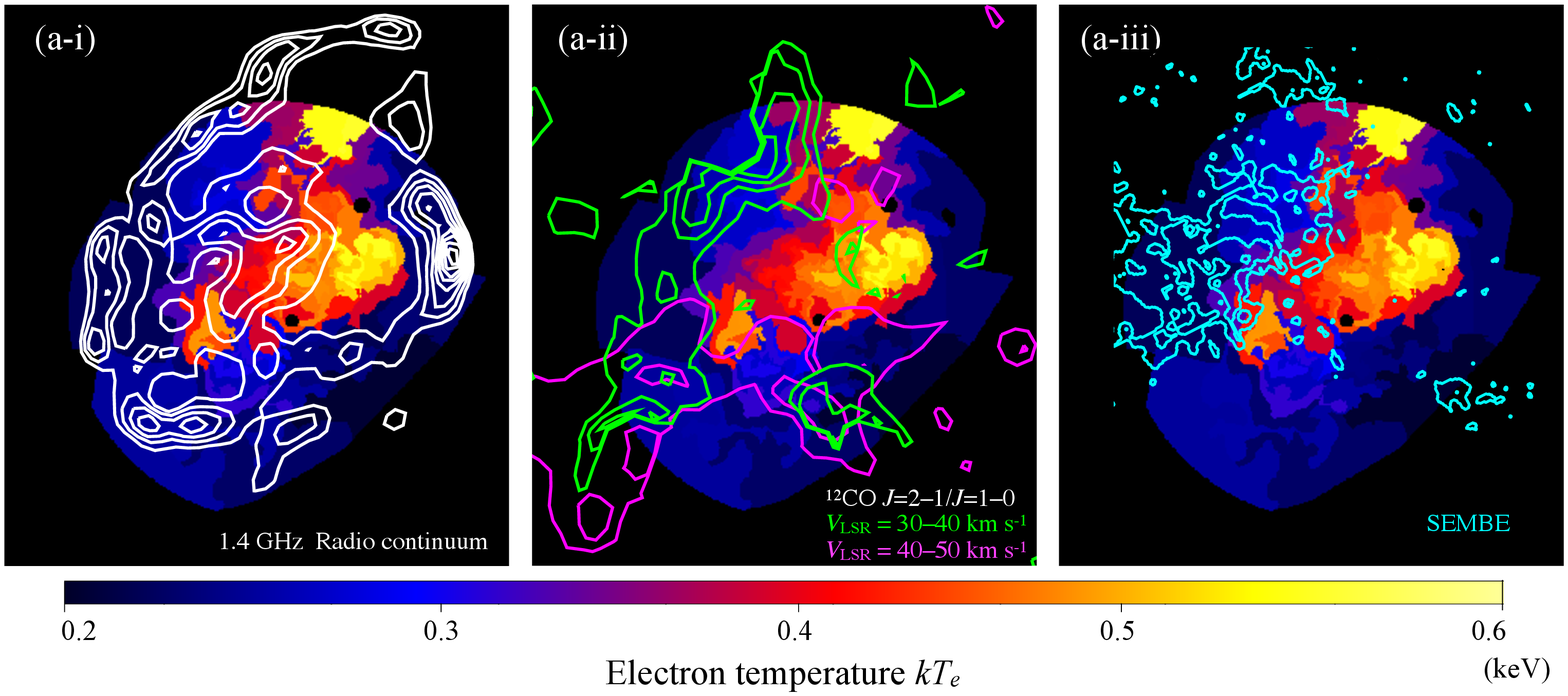} 
\end{center}
\vspace{-2mm}
\begin{center}
  \includegraphics[width=15cm]{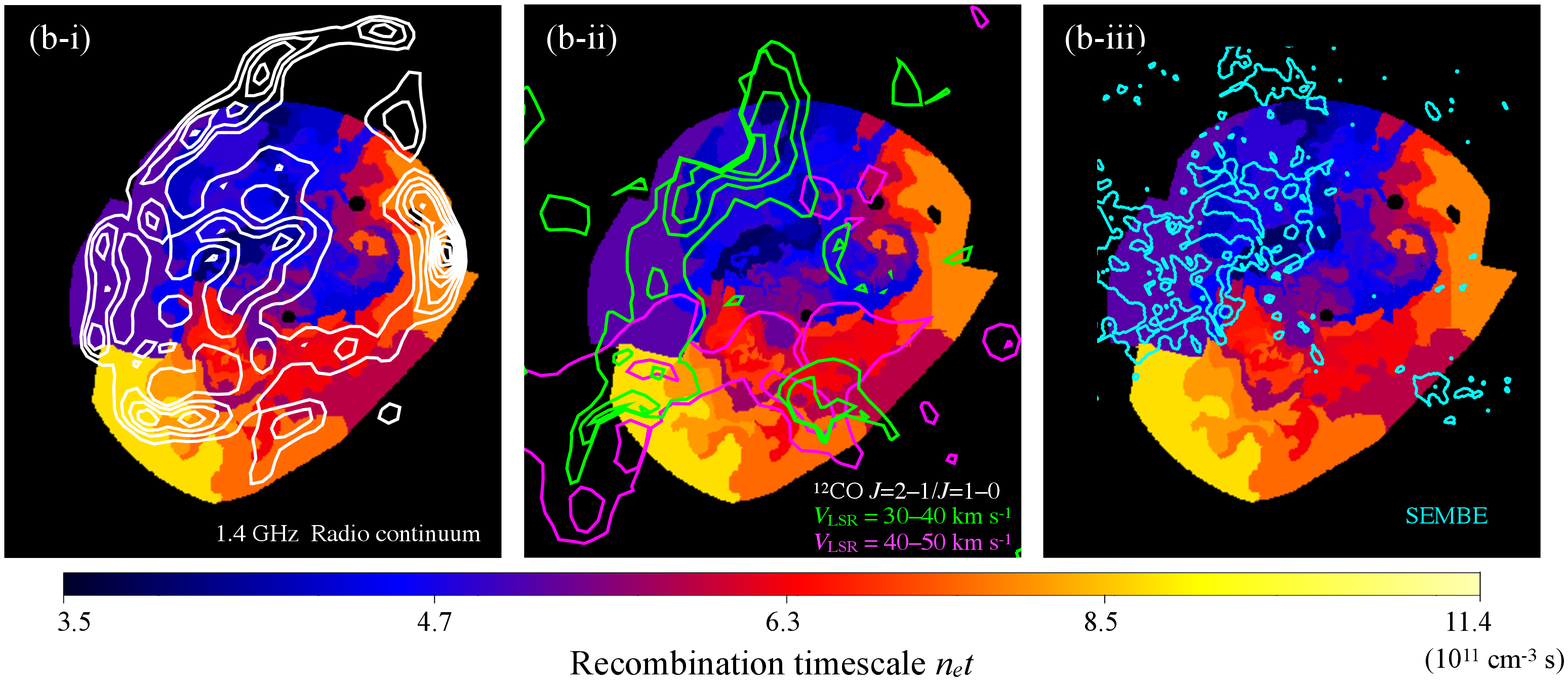} 
\end{center}
\vspace{-4mm}
\caption{
Maps presenting the distributions of (a) $kT_e$ and (b) $n_et$. 
The white contours in the panels (a-i) and (b-i) indicate the same radio continuum image as that in Figure~\ref{fig:Xray_image}(a). 
The green and magenta contours in panels (a-ii) and (b-ii) denote a $^{12}$CO($J=2$--$1$)-to-$^{12}$CO($J=1$--$0$) intensity ratio map drawn every 0.1 from 0.7 in $V_{\rm LSR} = 30$--$40~{\rm km}~{\rm s}^{-1}$ and  $V_{\rm LSR} = 40$--$50~{\rm km}~{\rm s}^{-1}$ taken from \cite{Yoshiike2013}. The velocity ranges correspond to that of the giant molecular cloud interacting with W44. 
In panels (a-iii) and (b-iii), the cyan contours denote the half-intensity line width diagram for the $^{12}$CO($J=1$--$0$) in $V_{\rm LSR} = 20$--$60~{\rm km}~{\rm s}^{-1}$ obtained with the Nobeyama 45 m radio telescope by \cite{Seta2004}.
Each contour represents 7, 10, and $13~{\rm km}~{\rm s}^{-1}$. 
}
\label{fig:kTe}
\end{figure*}

The X-ray absorption column densities ($N_{\rm H}$) obtained from the spectral fittings present a tool to probe the spatial distribution of the gas in front of W44. 
Figure \ref{fig:nH} shows a map of $N_{\rm H}$, where a significant spatial variation is clearly visible. 
The values, ranging from 1.3$\times10^{22}~{\rm cm^{-2}}$ to 2.8$\times10^{22}~{\rm cm^{-2}}$, are roughly consistent with the previous studies with {\it Suzaku} by \cite{Uchida2012} and \cite{Matsumura2018}, but the present map revealed a spatial distribution in much finer angular scales. 
We found that the X-ray absorption column densities are higher in the outer regions and are peaked in the northwestern rim. 
Based on CO line data, \cite{Seta2004} estimated the column density of foreground gas of  $\sim 1 \times10^{22}~{\rm cm^{-2}}$ and 
$\sim 2 \times10^{22}~{\rm cm^{-2}}$ in the inner and outer regions, respectively, which agree well with our X-ray results. 
Note that the ISM around W44 is dominated by molecular gas and that the contribution from atomic gas amounts to only $\sim 10\%$ of the total mass \citep{Yoshiike2013}. 
In the northwestern region with the highest X-ray absorption column density, it is known that a giant molecular cloud in the near side of W44 was hit by the SNR shock \citep{Seta2004}. 
When we select a velocity range of $40$--$50~{\rm km}~{\rm s}^{-1}$, which includes the most of the gas in the giant molecular cloud, the CO contours 
show almost a perfect match with the X-ray column density map (Figure~\ref{fig:nH}(b)). 
According to \cite{Seta2004}, the foreground gas in the northeastern rim has a column density of $N_{\rm H} \sim 3 \times 10^{22}~{\rm cm^{-2}}$, 
which is again consistent with the values estimated from our X-ray analysis.

\subsection{Physical Origin of RPs}
We now discuss the physical origin of the RPs in W44 based on spatial distributions of the parameters such as $kT_e$ and $n_et$, and 
their comparison with interacting gas distributions. 
Figure~\ref{fig:kTe}(a) shows a $kT_e$ map. 
In panel (a-ii), we overlaid a $^{12}$CO($J=2$--$1$)-to-$^{12}$CO($J=1$--$0$) intensity ratio map. 
The line ratio serves as a good indicator of shock-cloud interactions. 
A ratio well above the typical value in the unshocked part of the cloud, 0.6, indicates that the gas is shocked and/or heated \citep{Yoshiike2013}.
The electron temperature $kT_e$ tends to be lower at the locations where the $^{12}$CO($J=2$--$1$)-to-$^{12}$CO($J=1$--$0$) ratio is higher. 
Similar tendencies were found also in other MM SNRs with RPs, IC 443 and W28 \citep{Matsumura2017b, Okon2018}. 
Those authors claimed that the tendencies are most probably explained by thermal energy exchanges between the plasmas and clouds via thermal conduction \citep{Zhang2019}. 
Thus, our result on W44 would also be suggestive of significant thermal conduction between the X-ray emitting plasma and interacting gas. 

\begin{figure*}[ht]
\begin{center}
 \includegraphics[width=18cm]{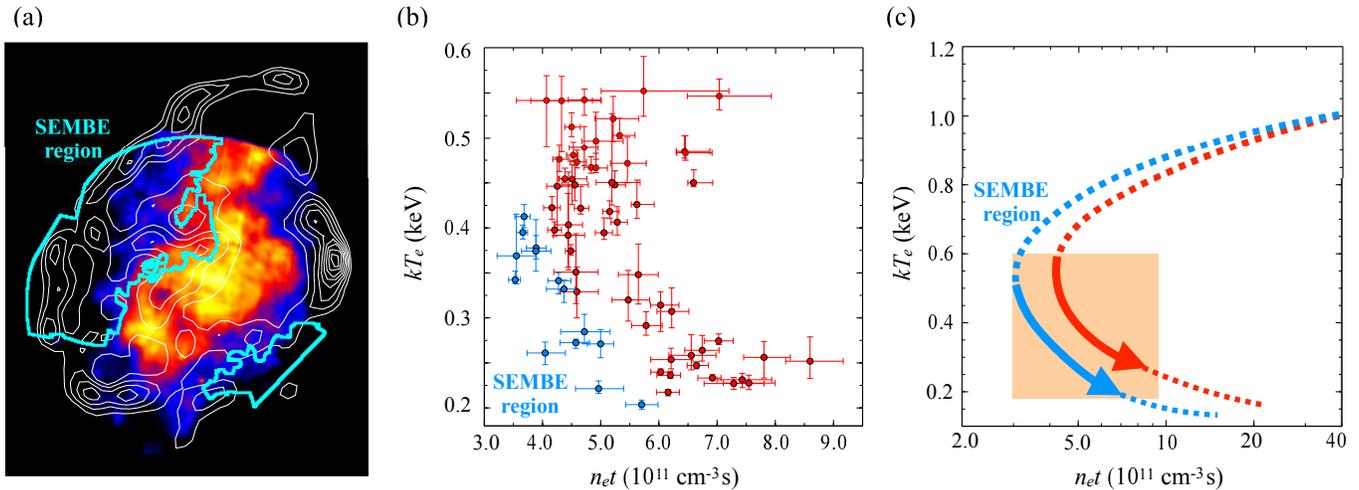} 
\end{center}
\vspace{-4mm}
\caption{
(a) Same as Figure~\ref{fig:Xray_image}(a) but with the definition of the SEMBE regions in cyan. (b) Relationship between $n_et$ and $kT_e$. The blue points are from the SEMBE regions whereas the red points are from the other regions. (c) Schematic trajectories of plasma during its evolution after a shock-cloud collision. The color scheme is the same as that of panel (b). The shaded area corresponds to the range plotted in panel (b). 
}
\label{fig:correlation}
\end{figure*}

Information from another parameter, $n_et$, whose spatial distribution is presented in Figure~\ref{fig:kTe}(b), has revealed far more convincing evidence for thermal conduction. 
In Figure~\ref{fig:correlation}, we plot $kT_e$ and $n_et$ derived for each sub-region.  
The data points are divided into two groups. 
The blue points come from regions where \cite{Seta2004} discovered $^{12}$CO($J=1$--$0$) lines broader than $\Delta V = 7~{\rm km}~{\rm s}^{-1}$ in full width at half maximum (see panels (a-iii) and (b-iii) of Figure~\ref{fig:kTe} and also Figure~\ref{fig:correlation}(a) for the locations), referred to as  spatially extended moderately broad emission (SEMBE) by \cite{Seta2004}. 
The red points, on the other hand, are from the other regions. 
The two groups are clearly separated from each other, 
and each of the two groups shows a clear anti-correlation between $kT_e$ and $n_et$. 

The result in Figure~\ref{fig:correlation}(b) can be well understood in a context of the thermal conduction scenario as follows. 
Let us assume that the X-ray emitting plasma initially had an ionization degree close to CIE ($n_et \sim 10^{12}~{\rm cm^{-3}s}$) and an electron temperature of $kT_{\rm init} \sim 1~{\rm keV}$ 
as we assumed in the spectral fittings in \S\ref{subsec:spec}. 
After the shock encountered the molecular cloud, the plasma was rapidly cooled due to thermal conduction. 
At this point, the plasma switched into an over-ionized state since the cooling proceeded in a timescale shorter than the recombination rate. 
Once the cooling rate became slower, recombination started to dominate to make the ionization degree gradually approach CIE. 
What we are currently observing in W44 would be emission from the plasma in this phase. 
Figure~\ref{fig:correlation}(c) presents schematic trajectories of the plasma on the $kT_e$-$n_et$ plane in the above scenario. 
The clear separation between the SEMBE and non-SEMBE regions suggests that the plasma was more efficiently cooled in the SEMBE regions. 

Although the nature of the SEMBE is not clear yet, a plausible interpretation would be that the SEMBE is emitted by unresolved dense clumps shocked and disturbed 
by the SNR shock \citep{Seta2004,Sashida2013}. 
Since those clumps are embedded in the hot X-ray emitting plasma, cloud evaporation would occur through thermal conduction between the plasma and the clumps, making 
the plasma in the SEMBE regions efficiently cooled. 
Cloud evaporation in SNRs are numerically studied with hydrodynamical simulations \cite[e.g.,][]{Zhou2011,Zhang2019} as well as with magneto-hydrodynamical 
simulations \cite[e.g.,][]{Orlando2008}.
According to the result by \cite{Zhang2019}, cloud evaporation plays a role in rapid cooling of hot plasma and thus also in over-ionization. 
\cite{Sashida2013} estimated the clumps in the SEMBE region have a size of $\ll 0.3$~pc.
The typical evaporation timescale of the clumps through thermal conduction can be evaluated as $t_{\rm evap} \approx 5.4\times10^{10} (n_e/1~{\rm cm^{-3}})(l/1~{\rm pc})^2 (kT_e/1~{\rm keV})^{-5/2}~{\rm s}\ll 10^{11}~{\rm s} $ \cite[e.g.,][]{Orlando2005} 
, assuming the plasma density $n_e = 1~{\rm cm}^{-3}$, the clump size $l \ll 0.3~{\rm pc}$, and the average plasma temperature $kT_e = 0.3~{\rm keV}$.
The evaporation timescale is sufficiently smaller than the timescale for a plasma to reach CIE ($t_{\rm CIE}\approx10^{12}\,(n_e/1~{\rm cm}^{-3})^{-1}$~s), and, therefore, 
cloud evaporation can make the plasma over-ionized.
ALMA would be able to resolve dense clumps of the SEMBE gas in W44 as pointed out by \cite{Sashida2013}. 
Spatially resolved spectroscopy in X-rays with angular scales similar to that of ALMA then should observationally reveal the process of cloud evaporation and resulting 
over-ionization.


In addition to the present result that has provided clear evidence for thermal conduction as the origin of the over-ionization in W44, 
recent studies suggest that the same mechanism seems responsible for RPs in other SNRs as well: G166.0+4.3 \citep{Matsumura2017a}, 
W28 \citep{Okon2018}, and CTB~1 \citep{Katsuragawa2018}.
A clear exception is W49B. 
Analyzing {\it NuSTAR} data of W49B, \cite{Yamaguchi2018} found a result opposite to this, a positive correlation between $kT_e$ and $n_et$. 
The correlation can be well explained by an efficient cooling by adiabatic expansion in low density areas.
The result led \cite{Yamaguchi2018} to conclude that adiabatic expansion plays a predominant role in over-ionizing the plasma in W49B. 
Then, what determines the main channel for rapid cooling? 
An important hint would be the young age of W49B \citep[1--6~kyr:][]{Pye1984,Smith1985b} in the category of MM SNRs containing RPs. 
Cooling through adiabatic expansion is expected to work efficiently in the early phase of SNR evolution when the blast wave breaks out of 
dense circumstellar matter into tenuous ISM \citep{Itoh1989,Shimizu2012}. 
On the contrary, thermal conduction can be effective only after the blast wave hits dense ambient ISM, and thus would be expected to be predominant at later stages. 
Based on hydrodynamical simulations,  \cite{Zhang2019} reached a similar conclusion that both thermal conduction and adiabatic expansion 
contribute to plasma cooling in SNRs and that the dominant channel can be changed as SNRs evolve. 
Systematic observational studies as well as their comparison with theoretical and numerical studies are necessary to 
further clarify the over-ionization process in SNRs.

\section{Conclusion}

We have performed spatially resolved spectroscopy of X-ray emission of W44 obtained with {\it XMM-Newton} deep observations, 
in order to clarify the physical origin of over-ionization of the X-ray emitting plasma. 
All spectra extracted from each region are well fitted with an RP model.
The X-ray absorption column densities well correlate with the distribution of foreground gas in the line-of-sight toward W44 as 
traced by radio line emissions reported in the literature.  
The obtained electron temperature $kT_e$ and the recombining timescale $n_et$ of RPs range from 0.18~keV to 0.60~keV and from $3  \times10^{11}~ {\rm cm}^{-3}~{\rm s}$ to $9 \times10^{11} ~{\rm cm}^{-3}~{\rm s}$, respectively. 
We have discovered that $kT_e$ is lower in the region where W44 is interacting with molecular clouds and that $kT_e$ and $n_et$ are 
negatively correlated. 
These findings indicate thermal conduction between X-ray emitting and the cold dense gas as the origin of the over-ionization. 
We have also found that $n_et$ is especially smaller in the regions with spatially extended moderately broad emissions of $^{12}$CO($J=1$--$0$) lines, which are considered to be emitted by clumpy gas shocked and disturbed by the SNR shock \citep{Seta2004,Sashida2013}.
This result can be explained by a rapid cooling of the plasma through evaporation of the clumpy gas. 
A comparison between our result on W44 and the result on W49B \citep{Yamaguchi2018} prompts us to consider that 
both thermal conduction and adiabatic cooling are possible channels of over-ionization in SNRs and that the dominant channel may change 
as SNRs evolve. 
Recent hydrodynamical simulations by \cite{Zhang2019} support this idea as well.

\acknowledgments

We deeply appreciate all the {\it XMM-Newton} and  NANTEN2  team members. 
This work is partially supported by JSPS/MEXT Scientific Research Grant Numbers JP19J14025 (H.O.), JP19H01936 (T.T.), JP25109004 (T.T. and T.G.T.), JJP19K03915 (H.U.), JP15H02090 (T.G.T.) , and ASI-INAF no. 2017-14-H.O (S.O., M.M., and F.B.)


\begin{thebibliography}{}

\bibitem[Arnaud(1996)]{Arnaud1996} Arnaud, K.~A.\ 1996, Astronomical Data Analysis Software and Systems V, 101, 17 


\bibitem[Claussen et al.(1997)]{Claussen1997} Claussen, M.~J., Frail, D.~A., Goss, W.~M., \& Gaume, R.~A.\ 1997, \apj, 489, 143 

\bibitem[Claussen et al.(1999)]{Claussen1999} Claussen, M.~J., Goss, W.~M., Frail, D.~A., \& Desai, K.\ 1999, \apj, 522, 349 



\bibitem[Foster et al.(2017)]{Foster2017} Foster, A.~R., Smith, R.~K., \& Brickhouse, N.~S.\ 2017, Atomic Processes in Plasmas (apip 2016), 190005.

\bibitem[Frail \& Mitchell(1998)]{Frail1998} Frail, D.~A., \& Mitchell, G.~F.\ 1998, \apj, 508, 690 


\bibitem[Greco et al.(2018)]{Greco2018} Greco, E., Miceli, M., Orlando, S., et al.\ 2018, \aap, 615, A157

\bibitem[Harrus et al.(1997)]{Harrus1997} Harrus, I.~M., Hughes, J.~P., Singh, K.~P., Koyama, K., \& Asaoka, I.\ 1997, \apj, 488, 781 


\bibitem[Itoh \& Masai(1989)]{Itoh1989} Itoh, H., \& Masai, K.\ 1989, \mnras, 236, 885 

\bibitem[Jones et al.(1993)]{Jones1993} Jones, L.~R., Smith, A., \& Angelini, L.\ 1993, \mnras, 265, 631 



\bibitem[Katsuragawa et al.(2018)]{Katsuragawa2018} Katsuragawa, M., Nakashima, S., Matsumura, H., et al.\ 2018, \pasj, 70, 110.

\bibitem[Kawasaki et al.(2002)]{Kawasaki2002} Kawasaki, M.~T., Ozaki, M., Nagase, F., et al.\ 2002, \apj, 572, 897 

\bibitem[Kawasaki et al.(2005)]{Kawasaki2005} Kawasaki, M., Ozaki, M., Nagase, F., Inoue, H., \& Petre, R.\ 2005, \apj, 631, 935 




\bibitem[Kuntz \& Snowden(2008)]{Kuntz2008} Kuntz, K.~D., \& Snowden, S.~L.\ 2008, \aap, 478, 575

\bibitem[Kushino et al.(2002)]{Kushino2002} Kushino, A., Ishisaki, Y., Morita, U., et al.\ 2002, \pasj, 54, 327 



\bibitem[Leahy, \& Tian(2007)]{Leahy2007} Leahy, D.~A., \& Tian, W.~W.\ 2007, \aap, 461, 1013

\bibitem[Lopez et al.(2013)]{Lopez2013} Lopez, L.~A., Pearson, S., Ramirez-Ruiz, E., et al.\ 2013, \apj, 777, 145 

\bibitem[Lumb et al.(2002)]{Lumb2002} Lumb, D.~H., Warwick, R.~S., Page, M., \& De Luca, A.\ 2002, \aap, 389, 93 


\bibitem[Matsumura et al.(2017a)]{Matsumura2017a} Matsumura, H., Uchida, H., Tanaka, T., et al.\ 2017, \pasj, 69, 30 

\bibitem[Matsumura et al.(2017b)]{Matsumura2017b} Matsumura, H., Tanaka, T., Uchida, H., Okon, H., \& Tsuru, T.~G.\ 2017, \apj, 851, 73 

\bibitem[Matsumura(2018)]{Matsumura2018} Matsumura, H. 2018, PhD thesis, Kyoto Univ.

\bibitem[Miceli et al.(2010)]{Miceli2010} Miceli, M., Bocchino, F., Decourchelle, A., et al.\ 2010, \aap, 514, L2



\bibitem[Nobukawa et al.(2018)]{Nobukawa2018} Nobukawa, K.~K., Nobukawa, M., Koyama, K., et al.\ 2018, \apj, 854, 87 

\bibitem[Okon et al.(2018)]{Okon2018} Okon, H., Uchida, H., Tanaka, T., Matsumura, H., \& Tsuru, T.~G.\ 2018, \pasj, 70, 35 

\bibitem[Ozawa et al.(2009)]{Ozawa2009} Ozawa, M., Koyama, K., Yamaguchi, H., Masai, K., \& Tamagawa, T.\ 2009, \apjl, 706, L71 


\bibitem[Orlando et al.(2005)]{Orlando2005} Orlando, S., Peres, G., Reale, F., et al.\ 2005, \aap, 444, 505 


\bibitem[Orlando et al.(2008)]{Orlando2008} Orlando, S., Bocchino, F., Reale, F., Peres, G., \& Pagano, P.\ 2008, \apj, 678, 274 


\bibitem[Petre et al.(2002)]{Petre2002} Petre, R., Kuntz, K.~D., \& Shelton, R.~L.\ 2002, \apj, 579, 404

\bibitem[Pye et al.(1984)]{Pye1984} Pye, J.~P., Becker, R.~H., Seward, F.~D., \& Thomas, N.\ 1984, \mnras, 207, 649

\bibitem[Ranasinghe \& Leahy(2018)]{Ranasinghe2018} Ranasinghe, S., \& Leahy, D.~A.\ 2018, \aj, 155, 204 


\bibitem[Rho \& Petre(1998)]{Rho1998} Rho, J., \& Petre, R.\ 1998, \apjl, 503, L167

\bibitem[Sanders(2006)]{Sanders2006} Sanders, J.~S.\ 2006, \mnras, 371, 829 


\bibitem[Sashida et al.(2013)]{Sashida2013} Sashida, T., Oka, T., Tanaka, K., et al.\ 2013, \apj, 774, 10 

\bibitem[Seta et al.(1998)]{Seta1998} Seta, M., Hasegawa, T., Dame, T.~M., et al.\ 1998, \apj, 505, 286 

\bibitem[Seta et al.(2004)]{Seta2004} Seta, M., Hasegawa, T., Sakamoto, S., et al.\ 2004, \aj, 127, 1098 

\bibitem[Sezer et al.(2019)]{Sezer2019} Sezer, A., Ergin, T., Yamazaki, R., et al.\ 2019, arXiv e-prints, arXiv:1907.01017

\bibitem[Shelton et al.(2004)]{Shelton2004} Shelton, R.~L., Kuntz, K.~D., \& Petre, R.\ 2004, \apj, 611, 906 

\bibitem[Shimizu et al.(2012)]{Shimizu2012} Shimizu, T., Masai, K., \& Koyama, K.\ 2012, \pasj, 64, 24 


\bibitem[Smith et al.(1985a)]{Smith1985a} Smith, A., Jones, L.~R., Peacock, A., \& Pye, J.~P.\ 1985, \apj, 296, 469 

\bibitem[Smith et al.(1985b)]{Smith1985b} Smith, A., Jones, L.~R., Watson, M.~G., et al.\ 1985, \mnras, 217, 99 


\bibitem[Str{\"u}der et al.(2001)]{Strder2001} Str{\"u}der, L., Briel, U., Dennerl, K., et al.\ 2001, \aap, 365, L18 

\bibitem[Turner et al.(2001)]{Turner2001} Turner, M.~J.~L., Abbey, A., Arnaud, M., et al.\ 2001, \aap, 365, L27 

\bibitem[Uchida et al.(2012)]{Uchida2012} Uchida, H., Koyama, K., Yamaguchi, H., et al.\ 2012, \pasj, 64, 141 

\bibitem[Uchida et al.(2015)]{Uchida2015} Uchida, H., Koyama, K., \& Yamaguchi, H.\ 2015, \apj, 808, 77 

\bibitem[Uchiyama et al.(2013)]{Uchiyama2013} Uchiyama, H., Nobukawa, M., Tsuru, T.~G., \& Koyama, K.\ 2013, \pasj, 65, 19 

\bibitem[Umemoto et al.(2017)]{Umemoto2017} Umemoto, T., Minamidani, T., Kuno, N., et al.\ 2017, \pasj, 69, 78



\bibitem[Wilms et al.(2000)]{Wilms2000} Wilms, J., Allen, A., \& McCray, R.\ 2000, \apj, 542, 914 

\bibitem[Wolszczan et al.(1991)]{Wolszczan1991} Wolszczan, A., Cordes, J.~M., \& Dewey, R.~J.\ 1991, \apjl, 372, L99 

\bibitem[Yamaguchi et al.(2009)]{Yamaguchi2009} Yamaguchi, H., Ozawa, M., Koyama, K., et al.\ 2009, \apjl, 705, L6 


\bibitem[Yamaguchi et al.(2018)]{Yamaguchi2018} \bibitem[Yamaguchi et al.(2018)]{2018ApJ...868L..35Y} Yamaguchi, H., Tanaka, T., Wik, D.~R., et al.\ 2018, \apjl, 868, L35 

\bibitem[Yoshiike et al.(2013)]{Yoshiike2013} Yoshiike, S., Fukuda, T., Sano, H., et al.\ 2013, \apj, 768, 179 


\bibitem[Zhang et al.(2019)]{Zhang2019} Zhang, G.-Y., Slavin, J.~D., Foster, A., et al.\ 2019, \apj, 875, 2 

\bibitem[Zhou et al.(2011)]{Zhou2011} Zhou, X., Miceli, M., Bocchino, F., Orlando, S., \& Chen, Y.\ 2011, \mnras, 415, 244 

\end{thebibliography}
\end{document}